\documentclass[twocolumn,preprintnumbers,amsmath,amssymb,prd,superscriptaddress,nofootinbib]{revtex4}

\usepackage{graphicx}
\usepackage{dcolumn}
\usepackage{bm}

\begin{document}


\title{
  Determination of the chiral condensate from 2+1-flavor lattice QCD
}

\newcommand{\Nagoya}{
  Department of Physics, Nagoya University, Nagoya 464-8602, Japan
}

\newcommand{\Taiwan}{
  Physics Department and Center for Theoretical Sciences, 
                National Taiwan University, Taipei, 10617, Taiwan  
}

\newcommand{\Tsukuba}{
  Graduate School of Pure and Applied Sciences, University of Tsukuba,
  Tsukuba 305-8571, Japan
}
\newcommand{\BNL}{
  Riken BNL Research Center, Brookhaven National Laboratory, Upton,
  NY11973, USA
}
\newcommand{\KEK}{
  KEK Theory Center,
  High Energy Accelerator Research Organization (KEK),
  Ibaraki 305-0801, Japan
}
\newcommand{\GUAS}{
  School of High Energy Accelerator Science,
  The Graduate University for Advanced Studies (Sokendai),
  Ibaraki 305-0801, Japan
}
\newcommand{\YITP}{
  Yukawa Institute for Theoretical Physics, 
  Kyoto University, Kyoto 606-8502, Japan
}
\newcommand{\Osaka}{
  Department of Physics, Osaka University,
  Toyonaka 560-0043, Japan
}

\author{H.~Fukaya}
\affiliation{\Nagoya}

\author{S.~Aoki}
\affiliation{\Tsukuba}


\author{S.~Hashimoto}
\affiliation{\KEK}
\affiliation{\GUAS}

\author{T.~Kaneko}
\affiliation{\KEK}
\affiliation{\GUAS}

\author{J.~Noaki}
\affiliation{\KEK}

\author{T.~Onogi}
\affiliation{\Osaka}

\author{N.~Yamada}
\affiliation{\KEK}
\affiliation{\GUAS}

\collaboration{JLQCD collaboration}
\noaffiliation


\begin{abstract} 
  We perform a precise calculation of the chiral condensate in QCD
  using lattice QCD with 2+1 flavors of dynamical overlap quarks. 
  Up and down quark masses cover a range between 
  3 and 100~MeV on a $16^3\times 48$ lattice 
  at a lattice spacing $\sim$ 0.11~fm.
  At the lightest sea quark mass, the finite volume system on the
  lattice is in the $\epsilon$ regime.
  By matching the low-lying eigenvalue spectrum of the Dirac operator
  with the prediction of chiral perturbation theory 
  at the next-to-leading order, we determine the chiral condensate 
 in 2+1-flavor QCD with strange quark mass fixed at its physical value as
  $\Sigma^{\overline{\mathrm{MS}}}(2\mathrm{~GeV})$ = 
  $[242(04)(^{+19}_{-18})\mathrm{~MeV}]^3$ where
  the errors are statistical and systematic, respectively.
\end{abstract}

\maketitle

Spontaneous breaking of chiral symmetry is one of the most fundamental 
properties of quantum chromodynamics (QCD),
as it produces the bulk of the hadron masses.
The symmetry breaking is indicated by a nonzero value of the chiral
condensate $\Sigma$, which is an expectation value of the scalar
density operator $\bar{q}q$. 
Despite its importance, 
calculation of $\Sigma$ remains a significant challenge,
even using the numerical simulation of QCD on the lattice,
due to both ultraviolet and infrared problems.

On the ultraviolet side, an additive renormalization of the 
scalar operator diverges as $\sim 1/a^3$ as the lattice spacing $a$
decreases, when the chiral symmetry is violated.
Even with exact chiral symmetry,
there exists a quadratic divergence proportional to the quark mass.
On the infrared side, since spontaneous symmetry breaking does not
occur at finite volume, 
the infinite volume limit has to be taken before going to the massless
limit. 
Therefore, careful study of the scaling in the chiral and infinite
volume limits is crucial to determine $\Sigma$.

Our previous work \cite{Fukaya:2007fb,Fukaya:2007yv} opened a new
possibility to overcome these difficulties by performing a lattice
QCD 
employing the overlap fermion formulation
\cite{Neuberger:1997fp,Neuberger:1998wv}, which preserves 
exact chiral symmetry at finite lattice spacings.
The ultraviolet problem is 
avoided
by using the spectrum of low-lying fermion modes. 
According to the Banks-Casher relation \cite{Banks:1979yr},
the spectral density $\rho(\lambda)$ of the Dirac operator at $\lambda=0$
is related to the chiral condensate as
$\Sigma=\pi\rho(0)$. 
At a large but finite volume $V$, 
chiral perturbation theory (ChPT) can be used to predict the volume
scaling of the near-zero modes, which is also equivalently described by 
the chiral random matrix theory 
\cite{Damgaard:1997ye, Wilke:1997gf, Akemann:1998ta, Damgaard:2000ah}.
By matching the theoretical prediction 
with the lattice data, the chiral condensate $\Sigma$ 
was determined at the leading order (LO) in the $\epsilon$ expansion 
(See also \cite{DeGrand:2006nv}).

This letter extends the previous work in several directions:
(i) Based on a new ChPT calculation by Damgaard and Fukaya
\cite{Damgaard:2008zs}, which is valid in the conventional $p$ regime
as well as in the $\epsilon$ regime, we use the lattice data 
at several values of sea quark masses.
(ii) The new formula consistently treats 
the next-to-leading order (NLO) effects in the
$p$ expansion and thus, the result of $\Sigma$ has the NLO accuracy
(A similar NLO analysis of the lattice data taken with the Wilson
fermion in the $p$ regime has been done recently \cite{Giusti:2008vb}.). 
(iii) The lattice data are newly generated including the effect of
strange quark, so that the result corresponds QCD in nature.
(iv) The finite volume scaling is confirmed using two volumes
$16^3\times 48$ and $24^3\times 48$.
With these new developments, the determination of $\Sigma$ is made
more precise and reliable.

The spectral density 
at a given topological charge
$Q$ is calculated within ChPT at NLO as \cite{Damgaard:2008zs}
\begin{equation}
  \label{eq:rho}
  \rho_Q(\lambda) = 
  \Sigma_{\rm eff}
  \hat{\rho}^\epsilon_Q(\lambda \Sigma_{\rm eff}V,\{m_{sea}\Sigma_{\rm eff}V\})
  + \rho^p(\lambda,\{m_{sea}\}),
\end{equation}
for an eigenvalue $\lambda$ of the Dirac operator.
Assuming the analyticity, $\rho_Q(\lambda)$ is obtained through the
 real part of the chiral condensate with a valence quark mass equal to an
 imaginary value $i\lambda$.
Here 
$\Sigma_{\mathrm{eff}}$ is an ``effective'' chiral condensate
of which definition is given below. 

The spectrum of the near-zero quark modes ($\lambda\sim 1/\Sigma V$)
is mainly affected by the zero-momentum pion modes.
In fact, the first term in (\ref{eq:rho}) is
the same as the spectral density at the 
leading order of the $\epsilon$ expansion
\cite{Damgaard:1997ye, Wilke:1997gf, Akemann:1998ta, Damgaard:2000ah} 
expressed 
as a function of
dimensionless combinations $\lambda\Sigma_{\rm eff} V$  and
$\{m_{sea}\Sigma_{\rm eff}V\}=\{m_1\Sigma_{\rm eff}V, \cdots,$
$m_{N_f}\Sigma_{\rm eff}V\}$ expressed by
\begin{eqnarray}
\hat{\rho}^\epsilon_Q(\zeta,\{\mu_{sea}\})
  \equiv C_2\frac{|\zeta|}{2\prod^{N_f}_f(\zeta^2 + \mu^2_f)}
  \frac{\det\tilde{\mathcal{B}}}{\det\mathcal{A}},
\end{eqnarray}
with $N_f\times N_f$ matrix $\mathcal{A}$ and 
$(N_f+2)\times(N_f+2)$ matrix $\tilde{\mathcal{B}}$ defined by
$\mathcal{A}_{ij}= \mu_i^{j-1}I_{Q+j-1}(\mu_i)$ and 
$\tilde{\mathcal{B}}_{1j} =  \zeta^{j-2}J_{Q+j-2}(\zeta)$,
$\tilde{\mathcal{B}}_{2j} =  \zeta^{j-1}J_{Q+j-1}(\zeta)$,
$\tilde{\mathcal{B}}_{ij} = (-\mu_{i-2})^{j-1}I_{Q+j-1}(\mu_{i-2})$ 
$(i\neq 1,2)$, respectively
($J_k$'s and $I_{l}$'s denote the (modified) Bessel functions.).
The phase factor $C_2$ is 1 for $N_f=2$ or 3.

The second term in (\ref{eq:rho}) is the NLO correction seen
in the ordinary $p$ expansion \cite{Osborn:1998qb}.
With the meson mass $M_{ij}^2\equiv (m_i+m_j)\Sigma/F^2$, which is 
made of either sea quark ($f$) or valence quark ($v$), it is
expressed as
\begin{eqnarray}
  \label{eq:rhop}
  \rho^p(\lambda,\{m_{sea}\}) &\equiv&
  -\frac{\Sigma}{\pi F^2}{\rm Re}\left[
  \sum^{N_f}_f (\bar{\Delta}(M^2_{fv})
  -\bar{\Delta}(M^2_{ff}/2))
\right.\nonumber\\&&\hspace{0.3in}\left.
  \left.-(\bar{G}(M^2_{vv})-\bar{G}(0))
  \right.\mbox{\huge $]$}\right|_{m_v=i\lambda}.
\end{eqnarray}
The function $\bar{\Delta}(M^2)$ contains the chiral logarithm,
$\bar{\Delta}(M^2)= 
\frac{M^2}{16\pi^2}\ln \frac{M^2}{\mu_{sub}^2}
+\bar{g}_1(M^2)$,
with $\bar{g}_1(M^2)$ representing the finite volume effect
\cite{Hasenfratz:1989pk}.
The subtraction scale $\mu_{sub}$ is set at 770~MeV in this work.
The other function $\bar{G}(M^2)$ has a double-pole contribution
due to the partial quenching.
The explicit forms of $\bar{g}_1(M^2)$ and $\bar{G}(M^2)$
are given in \cite{Damgaard:2008zs}
[In this Letter, we use a simplified notation. 
$\bar{G}(M^2)$ corresponds to $\bar{G}(0,M^2,M^2)$ in \cite{Damgaard:2008zs}.].
The effective chiral condensate $\Sigma_{\rm eff}$ 
in (\ref{eq:rho}) is given by
\begin{equation}
  \label{eq:Sigmaeff}
  \Sigma_{\rm eff} = 
  \Sigma \left[1-\frac{1}{F^2}\left(
      \sum^{N_f}_f \bar{\Delta}(\frac{M^2_{ff}}{2})-\bar{G}(0)
      -16L^r_6\sum^{N_f}_f M^2_{ff}\right)\right],
\end{equation}
where $L_6^r$ (renormalized at $\mu_{sub}$)
is one of the low-energy constants at NLO 
\cite{Gasser:1983yg}.

Numerical simulations of lattice QCD are performed using the
Iwasaki gauge action at $\beta=2.3$ including 2+1 flavors of
dynamical overlap quarks on a $16^3\times 48$ lattice.
The lattice spacing $a$ = 0.1075(7)~fm is determined from the heavy
quark potential with an input $r_0$ = 0.49~fm.
For the strange quark mass, we choose two different values 
$m_s$ = 0.080 and 0.100 in the lattice unit.
For the former, six values of up and down quark masses
$m_{ud}$ = 0.002, 0.015, 0.025, 0.035, 0.050, and 0.080 are taken.
For the latter, five values 
$m_{ud}$ = 0.015, 0.025, 0.035, 0.050, and 0.100 are used.
The smallest value $m_{ud}=0.002$
roughly corresponds to 3~MeV in the physical unit, 
with which pions are in
the $\epsilon$ regime. 
In order to investigate the finite volume scaling, we also simulate on a
$24^3 \times 48$ lattice at the same lattice spacing
with one choice of the sea quark masses $m_{ud}=0.025$ and $m_s=0.080$.

In the hybrid Monte Carlo (HMC) updates, the global topological
charge $Q$ of the gauge field is fixed to its initial value by
introducing extra (unphysical) Wilson fermions, 
which have a mass of cutoff order
\cite{Fukaya:2006vs}. 
In our main runs, we set $Q=0$.
We also simulate another sector of topological charge $Q=1$ 
at $m_{ud}$ = 0.015 and $m_s$ = 0.080.

We accumulate 2500 HMC trajectories for the main runs 
in the $p$ regime, 4750 (but the trajectory length is 0.5) 
for the $\epsilon$ regime lattice, 1800 for the $Q=1$ run,
and 1900 on the $24^3 \times 48$ lattice. 
Eighty pairs of low-lying eigenvalues 
of the massless
overlap operator $D$ are computed at every 5 
(or 10 in the $Q=1$ and $L=24$ runs) trajectories.
For the comparison with ChPT, every complex 
eigenvalue
$\lambda^{ov}$ is projected onto the imaginary axis as
$\lambda\equiv\mathrm{Im}\lambda^{ov}/(1-\mathrm{Re}\lambda^{ov}/(2m_0))$.
Here $m_0(=1.6)$ is a parameter to define the overlap-Dirac operator.
In the analysis, we consider positive $\lambda$ only.
The integrated autocorrelation time of the lowest $\lambda$ 
is measured as 6--24 trajectories depending on the simulation parameters.
The statistical error is estimated by the jackknife method
after binning data in every 100 trajectories.
Details of the numerical simulation will be reported elsewhere.

At each set of sea quark masses, the formula (\ref{eq:rho})
is described by two unknown quantities $\Sigma_{\rm eff}$ and $F$. 
Note that $\Sigma$ in (3) can be replaced by $\Sigma_{\mathrm{eff}}$ neglecting
higher order effects.
We first determine these parameters
from the lattice data of $\rho_Q(\lambda)$.
Roughly speaking, the height of $\rho_Q(\lambda)$ near $\lambda=0$
determines $\Sigma_{\rm eff}$ according to the Banks-Casher relation,
while the shape in the bulk region 
is controlled
by $F$, as far as $\lambda$ is in the region of convergence of the
chiral expansion.

\begin{figure*}[tbp]
  \centering
  \includegraphics[width=8cm]{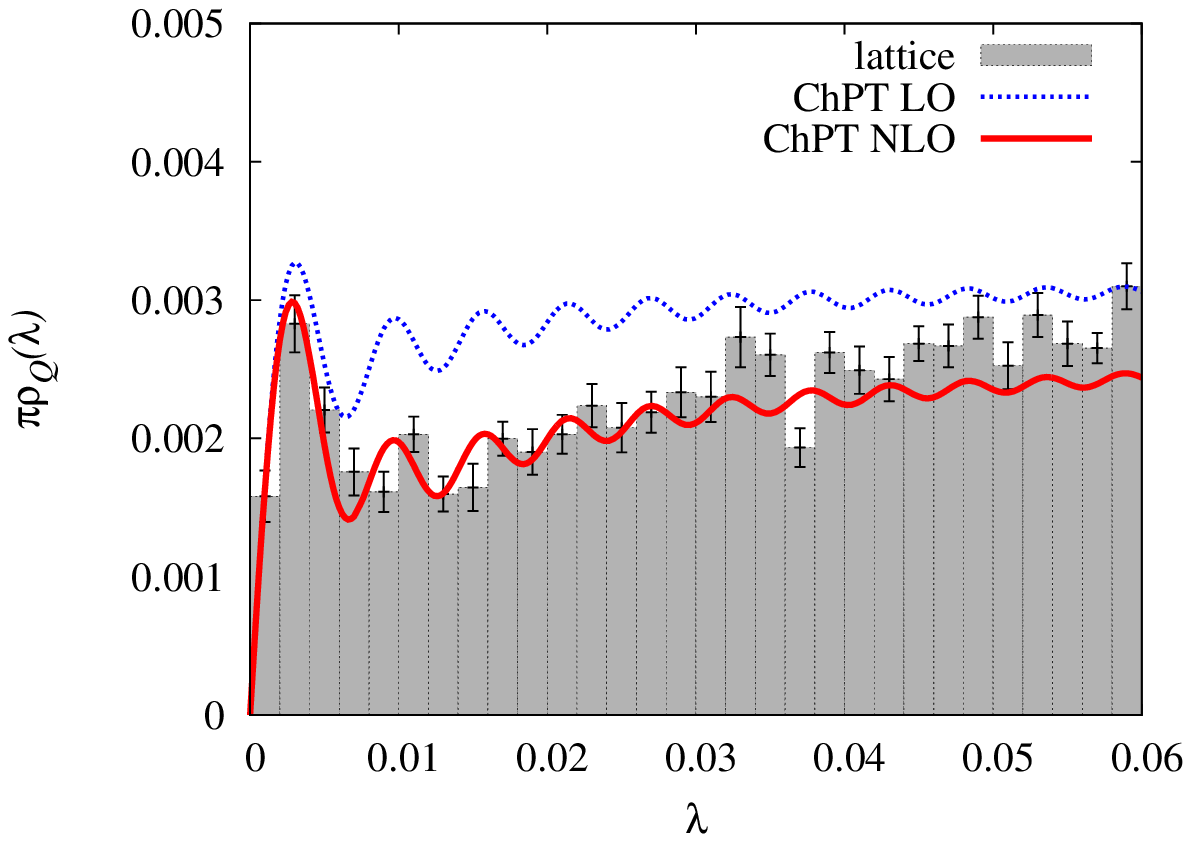}
  \includegraphics[width=8cm]{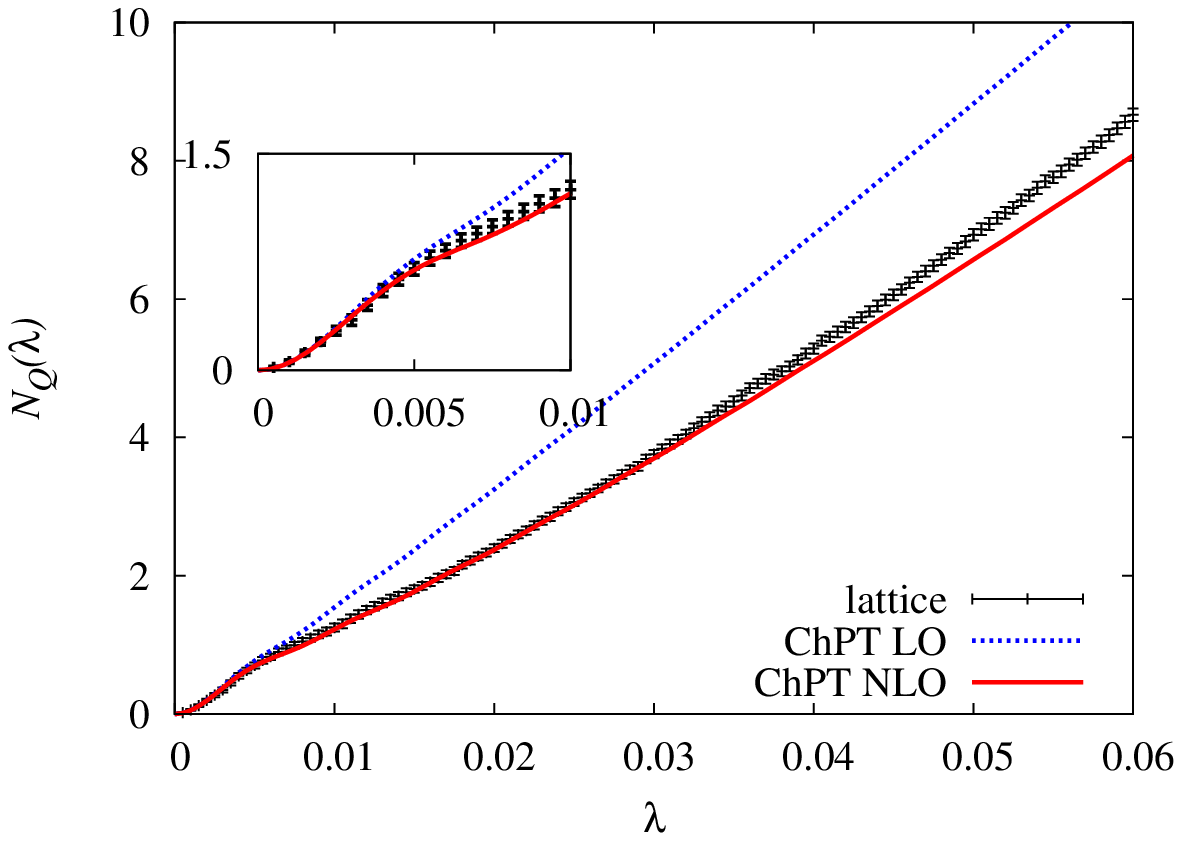}
  \caption{
    Spectral density $\pi\rho_Q(\lambda)$ (left) and mode number
    $N_Q(\lambda)$ (right) 
    of the Dirac operator at $m_{ud}$ = 0.015, $m_s$ = 0.080, and $Q=0$.
    The lattice result (given by histogram (left) or solid symbols 
    (right)) is compared with the ChPT formula (\ref{eq:rho}) 
    drawn by solid curves. 
    For comparison, the prediction 
    at the leading order of
    $\epsilon$ expansion (dashed curves) is also shown.
  }
  \label{fig:rho015}
\end{figure*}

Figure~\ref{fig:rho015} shows the spectral density $\rho_Q(\lambda)$
multiplied by $\pi$
(left panel) and the mode number below $\lambda$,
$N_Q(\lambda)\equiv V\int_0^\lambda d\lambda' \rho_Q(\lambda')$
(right), calculated at $m_{ud}$ = 0.015 and $m_s$ = 0.080.
The solid curve represents the ChPT result (\ref{eq:rho}) with
$\Sigma_{\rm eff}$ and $F$ determined from
$N_Q(\lambda)$ at two reference points
$\lambda$ = 0.004 ($\sim$ 7~MeV) and 0.017 ($\sim$ 30~MeV).
We observe that the formula (\ref{eq:rho}) describes the lattice data
well in the region below $\lambda\sim$ 0.03 ($\sim m_s/2$)
The result is stable within statistical error
under changes of the reference points in the range $\lambda<$ 0.03.
Beyond this value, higher order effects may become larger 
as suggested in the analysis of the pion mass and decay constant
\cite{Noaki:2008iy}. 
In the same figure, we also draw the first term of (\ref{eq:rho}).
Its discrepancy from the lattice data for 
$\lambda\gtrsim 0.01$ indicates 
that the second term $\rho^p(\lambda,\{m_{sea}\})$ in  (\ref{eq:rho})
is important for the consistency between QCD and ChPT.

\begin{figure}[tbp]
  \centering
  \includegraphics[width=8cm]{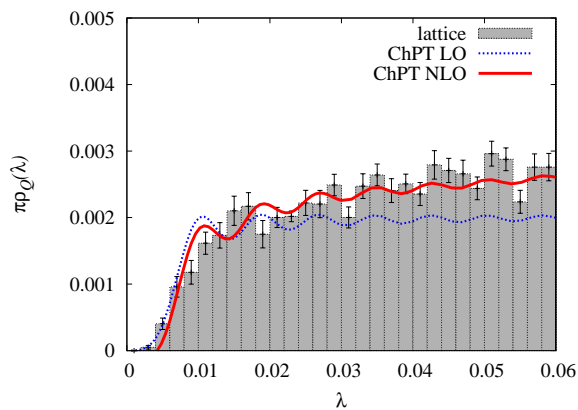}
  \caption{    
    Same as the left panel of Fig.~\ref{fig:rho015}, 
    but calculated on the $\epsilon$ regime lattice, 
    at $m_{ud}$ = 0.002 and $m_s$ = 0.080.
  }
  \label{fig:rho002}
\end{figure}

Results from the $\epsilon$ regime run are shown in Fig.~\ref{fig:rho002}.
$\Sigma_{\rm eff}$ and $F$ are determined from reference points 
$\lambda$ = 0.01 and 0.02.
We observe a good agreement between the data and 
NLO ChPT. 
The 
NLO correction is less significant than that in the $p$ regime, 
but still visible above $\lambda\sim$ 0.02.

\begin{table}[tbp]
  \centering
  \footnotesize
  \begin{tabular}{cccccc}
    \hline\hline
    & \multicolumn{2}{c}{$N_f=2+1$ formula} & 
    \multicolumn{2}{c}{$N_f=2$ formula} \\
    $m_{ud}$ & $\Sigma_{\rm eff}$ & $F$ & $\Sigma_{\rm eff}$ & $F$ &
    comment\\
    \hline
    0.002 & 0.00204(08) & 0.0465(100) & 0.00204(06) & 0.0423(49)\\
    0.015 & 0.00314(18) & 0.0536(15)& 0.00305(17) & 0.0551(16)\\
    0.015 & 0.00354(48) & 0.0521(25)& 0.00319(58) & 0.0558(62) & ($Q=1$)\\
    0.025 & 0.00333(18) & 0.0624(20) & 0.00326(18) & 0.0647(20)\\
    0.025 & 0.00306(07) & 0.0616(40) & 0.00304(07) & 0.0645(41) & ($L$=24) \\
    0.035 & 0.00404(39) & 0.0636(17) & 0.00393(36) & 0.0666(16)\\
    0.050 & 0.00423(22) & 0.0696(16) & 0.00413(21) & 0.0738(16)\\
    0.080 & 0.00453(23) & 0.0767(14) & 0.00444(22) & 0.0828(14)\\
    \hline
    0.015 & 0.00309(14) & 0.0564(19) & 0.00303(13) & 0.0578(19)\\ 
    0.025 & 0.00349(20) & 0.0622(17) & 0.00342(19) & 0.0642(17)\\ 
    0.035 & 0.00418(40) & 0.0647(14) & 0.00409(38) & 0.0673(14)\\ 
    0.050 & 0.00383(13) & 0.0713(16) & 0.00376(13) & 0.0747(16)\\ 
    0.100 & 0.00520(22) & 0.0835(14) & 0.00500(19) & 0.0924(16)\\
    \hline\hline
  \end{tabular}
  \caption{
    Numerical results for $\Sigma_{\rm eff}$ and $F$. 
    The upper half is the data at $m_s=0.080$ while the lower is at 0.100.
  }
  \label{tab:results}
\end{table}


The values of $\Sigma_{\rm eff}$ and $F$ 
are summarized in Table~\ref{tab:results} for all parameter choices.
We use the ChPT formulas of both $N_f=2$ and 2+1 cases.
The $N_f=2$ ChPT formula is understood as the leading
contribution in an expansion in terms of the large strange quark mass.
$\Sigma$ and $F$ in this framework depend on the
strange quark mass. 
The curves in Figs.~\ref{fig:rho015} and \ref{fig:rho002} are drawn
using the $N_f=2+1$ formula, but the difference from $N_f=2$ is hardly
visible in the range $\lambda<0.03$.
The numerical results of $\Sigma_{\rm eff}$ and $F$ are, in fact, 
insensitive to the choice of $N_f$ in the formula, except for $F$ in
the heavy mass region.
We also note that there is no significant difference of 
$\Sigma_{\rm eff}$ between $m_s$ = 0.080 and 0.100,
which confirms decoupling of the strange quark from the low-energy dynamics.

From the data in the non-trivial topological sector
$Q=1$,
we observe that the topological charge $Q$ largely affects 
the spectral density near $\lambda\simeq 0$,
but the values of $\Sigma_{\rm eff}$ and $F$ are consistent with those
at $Q=0$, as listed in Table~\ref{tab:results}.
The data 
at $L=24$ 
also show the expected scaling behavior from (\ref{eq:rho}).
Since the definition of $\Sigma_{\rm eff}$ (\ref{eq:Sigmaeff})
explicitly contains the lattice volume, the results from different
volumes cannot be compared directly.
After converting the $L=24$ lattice result $\Sigma_{\rm eff}$ =
0.00306(7) to that of $L=16$, we obtain 0.00341(18), which
is consistent with 0.00333(18) obtained on the $L=16$ lattice. 

\begin{figure}[tbp]
  \centering
  \includegraphics[width=8cm]{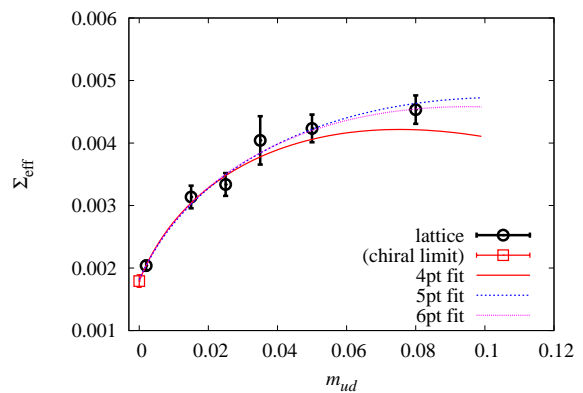}
  \caption{
    Three parameter fit of $\Sigma_{\rm eff}$
    to the $N_f=2+1$ ChPT.
  }
  \label{fig:SigmaefffitNf3}
\end{figure}

\begin{table}
  \centering
  \begin{tabular}{ll}
    \hline
    renormalization & $^{+1.2}_{-1.1}$ \% \\
    chiral fit & $^{+2.2}_{-0.7}$ \% \\
    finite volume & $^{+1.4}_{-0.0}$ \% \\
    finite $a$ & $\pm 7.4$ \% \\
    \hline
    total & $^{+7.9}_{-7.5}$ \% \\
    \hline
  \end{tabular}
  \caption{Systematic errors for
    $[\Sigma^{\rm phys}(2\mbox{~GeV})]^{1/3}$.
    The total error is obtained by adding each estimate by quadrature.
  } 
  \label{tab:sys}
\end{table}

Next, we analyze the sea quark mass dependence of $\Sigma_{\rm eff}$
from which 
$\Sigma$, $F$ and $L_6$ can be determined.
To see the convergence 
of the chiral expansion, we carry out
fits using four, five and six lightest data points as a function of
$m_{ud}$ 
with $m_s$ fixed at 0.080. 
The data points and fit curves of
the $N_f=2+1$ formula
are shown in Fig.~\ref{fig:SigmaefffitNf3}.
The curvature due to the chiral logarithm in (\ref{eq:Sigmaeff}) is
manifest.
The fit result for $\Sigma^{\rm phys}$, which is $\Sigma_{\rm eff}$ 
in the limit of $V=\infty$ and $m_{ud}=0$ while keeping
$m_s$ fixed at 0.08,
is stable under change of the fitting range.
Since we observe no sizable $m_s$ dependence (see Table~\ref{tab:results}),
$\Sigma^{\rm phys}$ can be considered as the one at the physical
strange quark mass.
From the five points fit, we obtain 
$\Sigma^{\rm phys}=0.00186(10)$, $F=0.0406(5)$ and $L_6^r=-0.00011(25)$
in the lattice unit, with $\chi^2/{\rm dof}=0.7$.



Since $F$ appears starting at the NLO correction in the formula, 
$m_{ud}$ dependence of the data given in Table~\ref{tab:results}
reflects the NNLO effects, 
which is beyond the scope of this work.
A naive linear extrapolation 
to the chiral limit yields
$F=0.0410(46)$, which roughly agrees with the value from the fit of
$\Sigma_{\rm eff}$. 

Our final result for the chiral condensate $\Sigma^{\rm phys}$,
in the limit of $m_{ud}=0$ and $m_s$ fixed at its physical value, is
\begin{equation}
  \label{eq:final}
  \Sigma^{\overline{\mathrm{MS}}}(2\mbox{~GeV}) 
  = [242(04)(^{+19}_{-18}) \mbox{~MeV}]^3,
\end{equation}
where the 
errors
are statistical and systematic, respectively.
The lattice scale $a$ = 0.1075(7)~fm is determined
from the heavy quark potential $r_0$ = 0.49~fm. 
We use the nonperturbatively calculated renormalization factor 
$1/Z_S(2\mbox{~GeV})$ = 0.806(12)($^{+24}_{-26}$)
\cite{Noaki:2009xi} to convert the result to the
$\overline{\mathrm{MS}}$ scheme at 2~GeV.

Possible systematic errors are listed in
Table~\ref{tab:sys}.
The error from chiral fit is estimated by taking variations of the
fitting range and the choice of $N_f$ in the ChPT formula.
Finite volume effect is estimated by taking the
difference between the data on $16^3\times 48$ and
$24^3\times 48$ lattices.
The discretization effect is hard to estimate within the calculation done
at a single lattice spacing, but partly reflected in the mismatch of
the lattice spacing obtained from different inputs: 
0.100(5)~fm from the pion decay constant \cite{JLQCD:2009sk}
and 0.109(2)~fm from 
the $\Omega$ baryon mass \cite{JLQCD:prep}. 
To be conservative, the maximum deviation from the central value
($\sim$~7.4\%) is added in both positive and negative directions
in Table~\ref{tab:sys}.

The chiral condensate obtained in this work (\ref{eq:final}) is
consistent with other determinations 
from the pseudoscalar meson mass, 
$\Sigma^{\overline{\mathrm{MS}}}(2\mbox{~GeV})=[257(14)\rm{MeV}]^3$
 \cite{JLQCD:2009sk}
and from the topological susceptibility,
 $\Sigma^{\overline{\mathrm{MS}}}(2\mbox{~GeV})=[249(4)\rm{MeV}]^3$
\cite{Chiu:2008kt,Chiu:2008jq}.
The former is obtained with the NNLO ChPT formula, while the latter
only uses the LO relation 
(and the errors do not contain the systematic effects).
Our result 
is also consistent with two-flavor
results in the previous works 
\cite{Fukaya:2007fb,Fukaya:2007yv,Giusti:2008vb, 
Noaki:2008iy,Aoki:2007pw,Fukaya:2007pn}.
Namely, there is no significant effect of the strange sea quark.

We also obtain $F$ = 74(1)(8)~MeV and $L^r_6$(770~MeV) = $-$0.00011(25)(11). 
Their systematic errors are estimated in a similar manner.

By the use of the eigenvalue density of the Dirac operator calculated
on the lattice, the chiral condensate is determined without
suffering from large subtraction of ultraviolet divergences. 
The dependence on the volume, topological charge and quark masses is
well described by ChPT at NLO in the region where both $\lambda$ and
$m_{ud}$ are smaller than $m_s/2$.

  HF thanks P.~H.~Damgaard for discussions.
  Numerical simulations are performed on the IBM System Blue Gene
  Solution at High Energy Accelerator Research Organization
  (KEK) under a support of its Large Scale Simulation
  Program (No. 09-05). 
  The work of HF was supported by the Global COE program of Nagoya
  University QFPU from MEXT of Japan.
  This work is supported in part by the Grant-in-Aid of the
  Japanese Ministry of Education 
  (No.~19540286, 20105001, 20105002, 20105003, 20105005, 20340047, 21684013).


\end{document}